# Pressure-controlled Structural Symmetry Transition in Layered InSe

Huimin Su [1†], Xuan Liu [2†], Chengrong Wei [1,7†], Junning Li [1], Zeyuan Sun [6], Qiye Liu [1], Xuefeng Zhou [1], Junhong Deng [2], Huan Yi [5], Qiaoyan Hao [5], Yusheng Zhao [1], Shanmin Wang [1], Li Huang [1], Shiwei Wu [6], Wenjing Zhang [5*], Guixin Li [2,3*], Jun-Feng Dai [1,3,4*]

1 Department of Physics, Southern University of Science and Technology, Shenzhen 518055, People's Republic of China
2 Department of Materials Science and Engineering, Southern University of Science and Technology, Shenzhen 518055, People's Republic of China
3 Shenzhen Institute for Quantum Science and Engineering, Southern University of Science and Technology, 518055, Shenzhen, People's Republic of China
4 Shenzhen Key Laboratory of Quantum Science and Engineering, Shenzhen 518055, People's Republic of China
5 International Collaborative Laboratory of 2D Materials for Optoelectronics Science and Technology, Key Laboratory of Optoelectronic Devices and Systems of Ministry of Education and Guangdong Province, and College of Optoelectronic Engineering, Shenzhen University, Shenzhen 518060, People's Republic of China
6 State Key Laboratory of Surface Physics, Key Laboratory of Micro and Nano Photonic Structures (MOE), and Department of Physics, Fudan University, Shanghai 200433, China
7 Physics Department, The University of Hong Kong, Pokfulam, Hong Kong

†These authors contributed equally to this work
*E-mail: daijf@sustech.edu.cn
        ligx@sustech.edu.cn
        wjzhang@szu.edu.cn

Abstract:

Structural symmetry of crystals plays important roles in physical properties of two-dimensional (2D) materials, particularly in the nonlinear optics regime. It has been a long-term exploration on the physical properties in 2D materials with various stacking structures, which correspond to different structural symmetries. Usually, the manipulation of rotational alignment between layers in 2D heterostructures has been realized at the synthetic stage through artificial stacking like assembling Lego bricks. However, the reconfigurable control of translational symmetry of crystalline structure is still challenging. High pressure, as a powerful external control knob, provides a very promising route to circumvent this constraint. Here, we experimentally demonstrate a pressure-controlled symmetry transition in layered InSe. The continuous and reversible evolution of structural symmetries can be in-situ monitored by using the polarization-resolved second harmonic generation (SHG) spectroscopy. As pressure changes, the reconfigurable symmetry transition of the SHG pattern from three-fold rotational symmetry to mirror symmetry was experimentally observed in a layered InSe samples and was successfully explained by the proposed interlayer-translation model. This opens new routes towards potential applications of manipulating crystal symmetry of 2D materials.

The structural symmetry of crystals shapes physical properties of materials, even playing a determinant role in electronic band structures and light-matter interactions[1,2]. For the two-dimensional (2D) materials, the interlayer-stacking configuration directly links to unusual electronic properties, such as long-lifetime interlayer excitons in 2D heterostructure[3,4]; Moiré bands in twisted bilayers[5,6]; superconductivity in magic-angle graphene[7]. With the aid of artificial stacking technique, a large number of 2D heterostructures or superlattices with fantastic properties can be fabricated by controlling the rotational alignment between layers, owing to weak van der Waals coupled interlayer bonding. Another equally important parameter, namely the relative translation between layers, which correspond to translational symmetry of crystalline structure of 2D materials, also plays an important role in exploring emergent physical properties. However, according to density functional theory (DFT) calculation[8], the structures with unusual interlayer translation possess a relatively high total energy, indicating that they are not stable in constituent layers. Theoretically, such crystalline translation may be achieved under loading hydrostatic pressure. As the interlayer spacing decreases with increasing pressure, the enhanced interlayer interaction may cause a relative sliding between adjacent layers, which contribute to symmetry transition[9,10]. However, this kind of structural translation in 2D crystals has not yet been observed, partially due to the lack of an in-situ method to monitor and validate this change.

Layer structured InSe shows very high nonlinear optical coefficients in the infrared (IR) range[11,12]. Its monolayer has a honeycomb lattice and is composed of covalently bonded Se–In–In–Se atomic planes[13]. It crystallizes to a non-centrosymmetric $D_{3h}^1$ space group, which would lead to a non-zero SHG with the normally incident fundamental wave (FW) along its c-axis. For $\varepsilon$ −InSe (see Fig. 1a), the rhombohedral stacking order indicates that inversion symmetry breaking still exists in multilayer samples. Also noted is that the layered InSe possesses both broken inversion symmetry and threefold rotational symmetry. In this work, the pressure-induced reversible structural translation was demonstrated successfully in layered InSe with high nonlinear optical coefficients, thanks to the sensitive second-order nonlinear optical spectroscopy.

Here we report that by applying hydrostatic pressure to layered InSe, a symmetry transition from three-fold rotational symmetry to mirror symmetry was observed in the orientation-dependent SHG pattern. The theoretical modeling indicates that relative sliding between layers plays an important role. More importantly, the structural translation and the nonlinear optical properties could be restored to the initial states when pressure is released. Our study demonstrates a practical way towards reconfigurable manipulation of structure symmetry of 2D materials. Under various pressure conditions, the effective nonlinear susceptibility of 2D crystal can be retrieved and then the symmetry transition can be indirectly quantified.

High quality InSe single crystal was synthesized by modified Bridgman method, whose structure was experimentally confirmed to be the $\varepsilon$-phase. The multilayer samples were mechanically exfoliated from InSe crystal onto a Si substrate with 300-nm-thick $SiO_2$. To avoid possible reactions with chemical species in air, such as oxygen, samples were mounted in a nitrogen-filled box with a 1-inch-diameter optical window for optical measurements. For SHG measurements in the layered InSe under hydrostatic pressure, the samples were exfoliated directly onto the surface of one diamond, which serves as a part of the diamond anvil cell (DAC) (see Methods). The whole sample preparation process was completed in a glovebox filled with dry nitrogen to guarantee the uncontaminated surface. Both linearly and circularly polarization-resolved SHG measurements were performed in a back-reflection geometry using a normally incident FW (see Methods). Fig. 1b presents the excitation-power dependence of the SHG intensity at a fundamental wavelength of 1060 nm. We observed the SHG signal at 530 nm with a near-quadratic (slope = 2) dependence on the excitation power (red line in Fig. 1b), which agrees well with the expectation for SHG process[14]. Fig.1c shows the angle-resolved SHG spectrum, where SHG signals with polarization parallel (blue circles) to that of FW exhibit a sinusoidal function dependence on the in-plane polarization angle 3θ of FW, indicating three-fold rotational symmetry of the layered InSe. Fig. 1d shows the polarization-resolved SHG mapping conducted in a ~10 nm thick layered InSe, where the polarization of FW and SHG signals are parallel. The uniformity of SHG intensities reflects the negligible microscopic crystalline grains and domains of the samples[15,16].

It is known that for a natural or artificial optical crystal with three-fold rotational symmetry, the polarization state of SHG follows specific selection rule, i.e., SHG has opposite circular polarization to that of FW propagating along the rotational axis of the crystal (Fig. 2a)[17-20]. In the circularly polarization-resolved SHG measurement, layered InSe was excited by the left- (LCP) and right- (RCP) circularly polarized FW at the wavelength of 1060 nm, respectively. In this case, the interband absorption is avoided, as the excitation energy (1.17 eV) is less than the band gap of InSe at 1.27 eV (Fig. S1). As shown in Fig. 2b, it is observed that under the pumping of the circularly polarized FW, the SHG wave at the wavelength of 530 nm with opposite circular polarization is much stronger than that with same polarization state as the FW. The purity of circular polarization state of SHG is evaluated by using the ellipticity $P = \frac{I_{\sigma+} - I_{\sigma-}}{I_{\sigma+} + I_{\sigma-}}$, where $I_{\sigma+}(I_{\sigma-})$ is the intensity of right (left)-circular polarization component of SHG. For a typical sample shown in Fig. 2b, the averaged SHG ellipticity is around 0.93 and -0.91 for FW with LCP and RCP states, respectively. It is noted that the ellipticity is very uniform among the entire sample, despite different thicknesses (Fig. S2). It is expected from unaffected stacking order in $\varepsilon$-InSe with different thicknesses. Meanwhile, the non-unity ellipticity might result from the imperfection of the measurement system. A similar phenomenon of circular SHG was observed in other 2D materials with threefold rotational symmetry such as monolayer $WSe_2$[21,22].

Then, we dynamically tune the structural symmetry by using hydrostatic pressure and monitor any transition or deformation of the structures via polarization-resolved SHG spectroscopy. Fig. 3a shows the schematic of the polarization-resolved SHG setup under pressure. For 2D layered InSe with nanometer thickness, this pressure is equivalent to an out-of-plane pressure due to a large ratio of surface to volume. Previous studies justified that bulk InSe undergoes a first order phase transition to the NaCl-type structure under pressure above 10 GPa[23]. To avoid this kind of phase transition, the pressure in this experiment is limited to be below 8.2 GPa. From the reflection color of the samples, the degradation of the layered crystal has not been observed within the range of the applied pressures. For measurements under all pressures, we cannot observe any obvious rotation of the six-petal of SHG pattern, where the symmetry plane of the crystal has a fixed angle of $\theta = 45°$ with respect to the X-axis of the laboratory coordinate[24].

As shown in Fig. 3b, the ellipticity of circular SHG exhibits a series of changes as pressure increases to 8.2 GPa. Below 3.0 GPa, the ellipticity remains unchanged at around 90%. Above that value, it decreases gradually as pressure increases and approaches zero for pressure above 6 GPa (see the black squares in Fig. 3b). It indicates the breaking of the threefold rotational symmetry of crystal structure as pressure increases. To figure out what happened, linearly polarization-resolved SHG spectroscopy was conducted to determinate the evolution of structural symmetry under the same pressure range. Fig. 3c shows the evolution of linear SHG patterns under various pressure conditions, where the red circles and black squares represent the intensities of parallel ($I_\parallel$) and perpendicular ($I_\perp$) components of SHG waves, respectively. If no pressure is applied, the two components of SHG intensity exhibit a six-petal pattern with the same rotational symmetry but a phase difference of 60 degree, revealing the threefold rotational symmetry of layered InSe. As pressure increases, the two six-petal patterns ($I_\perp$ and $I_\parallel$) are gradually deformed, but the trends are different. For the $I_\perp$ component, four of the six petals are gradually suppressed and the pattern trends to a dumbbell-like pattern at 8 GPa. Meanwhile, for the $I_\parallel$ component, the pattern gradually extends along the direction perpendicular to 45 degree axis of symmetry and changes to an apple-like shape at 8 GPa (details are shown in Fig. S3). Under each pressure, SHG pattern exhibits the similar shape at different points on the sample (it is not shown). The evolution of azimuthal polarization pattern revealed by linearly polarization-resolved SHG spectra shows a symmetry transition from the threefold rotational symmetry to the mirror symmetry, which is consistent with the change of ellipticity of SHG under different pressures. Importantly, as the pressure is reduced, the six-petal pattern and corresponding ellipticity were gradually restored to their initial states as shown by the last three patterns in Fig. 3c and red circles in Fig. 3b, respectively. This re-configurability also justifies that sample degradation is negligible in the experiment. It is interesting that the pattern evolution is unchanged if the pressure is released and reapplied once again. But for two samples with the similar initial orientation, distortion direction can be completely different, where maximum SHG intensity is at 10 degree for S1 and 70 degree for S2, respectively (see Fig. S4 in the supplementary materials). We speculate that the difference in distortion direction is related to the initial direction of pressure loading. Further experiments are needed to clarify this difference. So far, we have achieved the continuous tunability of interlayer interaction strength through an out-of-plane pressure and observed the evolution of polarized SHG emission under the pressure.

To better understand the pressure-induced evolution of the SHG patterns in Fig. 3c, we developed a model to calculate the polarization and intensity of SHG (see details in the supplementary materials). If no pressure is applied onto the sample, we can define the relative value of the non-zero elements of $\chi^{(2)}_{D_{3h}}$, which are normalized with $\chi^{(2)}_{yyy} = -\chi^{(2)}_{yxx} = -\chi^{(2)}_{xxy} = -\chi^{(2)}_{xyx} = 1$. As shown in Fig. 4a-4d, with increasing pressure along the c-axis of InSe crystal, we assume that the crystal has a tiny displacement between layers occurs, exhibiting a deformation from perfect threefold rotational symmetry. To further simplify the calculation, we only consider the changes of the four nonzero second-order susceptibilities mentioned above. By fitting the experimental results in Fig. 3c, we successfully retrieved the effective susceptibilities $\chi^{(2)}$ under different pressures (Table I in the supplementary materials). It was found that the calculated polarization properties of SHG agree very well with the experimental results. Under the pressure of our experiment, $\chi'^{(2)}_{yxx}$ in the local coordinate does not change too much. By contrast, $\chi'^{(2)}_{yyy}$ and $\chi'^{(2)}_{xxy} = \chi'^{(2)}_{xyx}$ are highly sensitive to the deformation of the crystal and phase angle in two terms gradually increases and decreases with increasing pressure, respectively. More interestingly, although the absolute value of effective $\chi^{(2)}$ does not deviate too far from its initial value, the polarization-resolved SHG signal indeed evolves quickly from a threefold symmetric pattern to a mirror symmetric pattern. This means that the polarization-resolved SHG spectroscopy is a powerful tool to investigate the structural symmetry in 2D materials when the high pressure is applied as a noninvasive technique. In contrast, traditional methods in high-pressure research such as XRD and Raman techniques[25,26], which are sensitive to layer distance, cannot give any information about the relative sliding or dislocation between layers.

With hydrostatic pressure applied, ε-InSe may undergo several structure-evolutions, such as adjacent layer sliding, interlayer rotation, or intralayer In-Se covalent bond variation and even phase transition into the metastable phases. The existence of a rock-salt (RS) cubic phase[23] (space group $D_{4h}$) and a monoclinic (MC) structure (space group $C_{2h}$)[27,28] have been reported for InSe at the low pressure region, which is metastable in ambient conditions[29]. Crystals with either the $D_{4h}$ or $C_{2h}$ structure give no SHG emission with normal incident laser beam[30], which would lead to a steep discontinuity on the pressure-dependent SHG curve at the critical pressure. However, such discontinuity were absent in our experiments at the pressure range of 0-8 GPa, and the total SHG signal remains on the same order of magnitude at higher pressures (see Fig. S5 in the supplementary materials), suggesting no phase transition among these metastable phases during the process. On the other hand, the orientation rotation of adjacent layers under pressure can only change the orientation of the symmetry axes in the SHG pattern but not the symmetry of the pattern[24]. These imply that the adjacent layer sliding of ε-InSe occurs under pressure, which is the primary reason for the observed SHG pattern variation.

Microscopically, the relative sliding between adjacent layers is mainly due to changed interactions between on-top atoms as interlayer distance decreases[9,10]. Based on the density functional theory (DFT) calculation, we got the potential energy surface for various crystal structures achieved by sliding between layers along a direction of Se-In dimer and its perpendicular direction in bilayer $\epsilon$−InSe (Fig. S10-S11). Our results indicate the relative sliding along a direction of Se-In dimer in x-y plane has a perfect mirror symmetrical structure and a relative low potential barrier, which maybe contribute to the distortion of SHG patterns under high pressure. However, for 20 nm thick $\epsilon$−InSe with a repeat unit of two layers, it contains at least 25 layers with an interlayer distance of 0.8 nm[13,31]. As interlayer interaction increases, the form of relative sliding among all the layers is very complicated, where sliding parallel and perpendicular to the direction of Se-In dimer will occur simultaneously among different layers. Therefore, it is difficult to determine which specific sliding form contributes to distortion of SHG spectra in multilayer InSe under all the pressure. As shown in Fig. 3c and Fig. S3, the azimuth polarization pattern above 3 GPa shows that it is no perfect mirror symmetry, which also indicates complicacy of atomic-scale sliding. Further study on pressure-tunable SHG response in monolayer and bilayer 2D materials is needed to clarify the specific physical processes under hydrostatic pressure.

In summary, we demonstrate a reconfigurable structural symmetry transition of layered InSe, i.e., from the threefold rotational symmetry to the mirror symmetry, by applying the out-of-plane pressures, and vice versa. This kind of phase transition manifested by the change of structural symmetries can be directly characterized by using the polarization-resolved SHG microscopy. The nonlinear optical calculations and DFT results further verify that the relative sliding between layers plays a very important role in the evolution of polarizations of SHG waves. The continuous tunability of the structural translation symmetry under hydrostatic pressure presented in this work can be extended to other 2D materials and probed in a noninvasive way thanks to the powerful SHG technique.

**Methods**
**1. High-pressure experiment.** The high-pressure experiments were conducted by using a diamond anvil cell (DAC), whose pressure can be up to 10 GPa. Diamond culets are cut parallel to the 100 planes to reduce polarization dependent response of transverse light. The chamber for loading samples has a cylinder-shaped hole with a diameter of 300 μm and height of 30 μm, which was fabricated in the center of a stainless-steel gasket. To obtain hydrostatic pressure conditions, the admixed methanol and ethanol in the molar ratio 4: 1 as the pressure-transmitting medium, was evenly filled into the chamber. A few ruby balls were also loaded into the sample chamber to serve as the internal pressure standard. In the end, the face of two diamond-anvils is exactly aligned with each other, in order to achieve a uniform hydrostatic pressure on the sample.
**2. SHG measurement.** The SHG measurement was performed in our homemade system with a back-reflection geometry. The fundamental wave with a central wavelength of 1060 nm was generated from a Ti-Sapphire oscillator (Chameleon Ultra

II) with an 80 MHz repetition and 150 fs pulse width. The pump beam was focused to a spot of around 2 μm on the sample by using a 50× objective. The reflected SHG signal was collected by the same objective and then was fed into a spectrometer equipped with a thermoelectric cooled charged-coupled device (CCD). Two shortpass filters with a central wavelength of 750 nm were used to block the fundamental wave. For linearly polarization-resolved SHG measurement, the polarization direction of the fundamental wave is tuned by a 1/2 waveplate with a wavelength range of 310-1100 nm. The parallel and perpendicular components of reflected SHG signals pass through the same 1/2 waveplate, a non-polarized beamsplitter and a displacer, and then were focused to two spots on the slit of the spectrometer. By rotating the fast axis of the 1/2 waveplate, the intensity of two orthogonal polarization components of SHG as a function of crystal orientation was measured. For circularly polarization-resolved SHG measurement, the 1/2 waveplate was replaced by a 1/4 waveplate. Thus, the polarization of fundamental wave was converted to a circular polarization. Two circularly polarized components of SHG signals were converted to two orthogonally linearly polarized components by the same 1/4 waveplate, and then focused onto two spots after passing through the displacer and finally recorded by the CCD spectrometer. The ellipticity is then obtained according to the following equation: $P = \frac{I_{\sigma+} - I_{\sigma-}}{I_{\sigma+} + I_{\sigma-}}$, where $I_{\sigma+}$ ($I_{\sigma-}$) is the intensity of right-circular (left-circular) polarization components of SHG intensity.

**Acknowledgements**
We would like to thank Prof. Xiao-Dong Cui and Prof. Kedong Wang for helpful discussions. J. F. acknowledges the support from the National Natural Science Foundation of China (11204184), Special Funds for the Development of Strategic Emerging Industries in Shenzhen (no. JCYJ20160613160524999), Science, Technology and Innovation Commission of Shenzhen Municipality (Grant No.ZDSYS20170303165926217), and the Guangdong Natural Science Foundation (2017A030313023). G. L. acknowledges the support from the National Natural Science Foundation of China (11774145), Guangdong Innovative & Entrepreneurial Research Team Program (2017ZT07C071), Applied Science and Technology Project of Guangdong Science and Technology Department (2017B090918001) and Natural Science Foundation of Shenzhen Innovation Committee (JCYJ20170412153113701). H. M. acknowledges the support from the National Natural Science Foundation of China (11604139). S. W. acknowledges the supported from the Shenzhen Peacock Plan (No. KQTD2016053019134356) and Guangdong Innovative & Entrepreneurial Research Team Program (No. 2016ZT06C279). Y. Z. acknowledges the Shenzhen Development and Reform Commission Foundation for Novel Nano-Material Sciences.

**Author contributions**
J.D. and G.L. conceived the project. J.D., H.S., C.W. and Q.L. designed the experiments. H.S. and C. W. performed the experiments. H. Y., Q. H. and W.Z. provided the samples. X. Z., S. W. and Y. Z. provided the DAV technique. X.L., J.L., J.D., L.H., S. W. and

G.L. provided the theoretical supports. All authors discussed the results and co-wrote the paper.

**Additional information**
The authors declare no competing financial interests.

**Figures and captions**

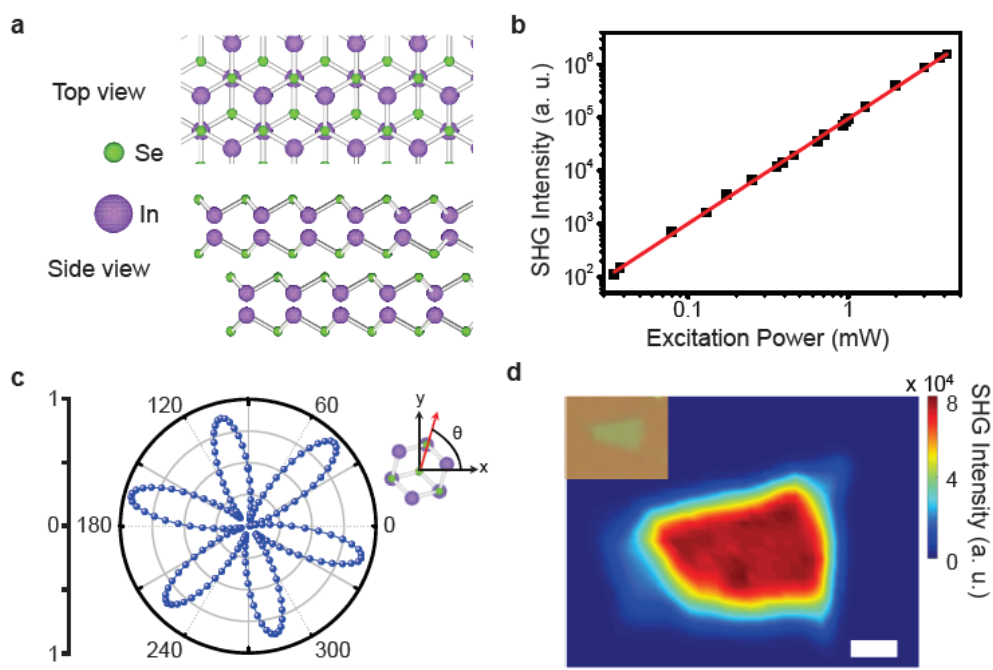

**Figure 1 SHG property of layered InSe.** (a) Side view and top view of bi-layer crystal structures, which serves as a repeat period in $\epsilon-$InSe. Purple and green spheres correspond to indium (In) and selenium (Se) atoms, respectively. (b) SHG intensity as a function of excitation power (black squares). The fitting curve (red line) shows that the intensity of SHG is the square of that of the fundamental wave. (c) The intensity of parallel component of SHG signal as a function of θ, namely the angle between the x-axis of laboratory coordinate and the crystalline orientation. It exhibits the expected $\sin^2(3\theta)$ dependence. (d) Polarization-resolved SHG mapping in a homogeneous layered InSe. The inset shows the optical image of the layered sample. Scale bar: 5 $\mu m$.

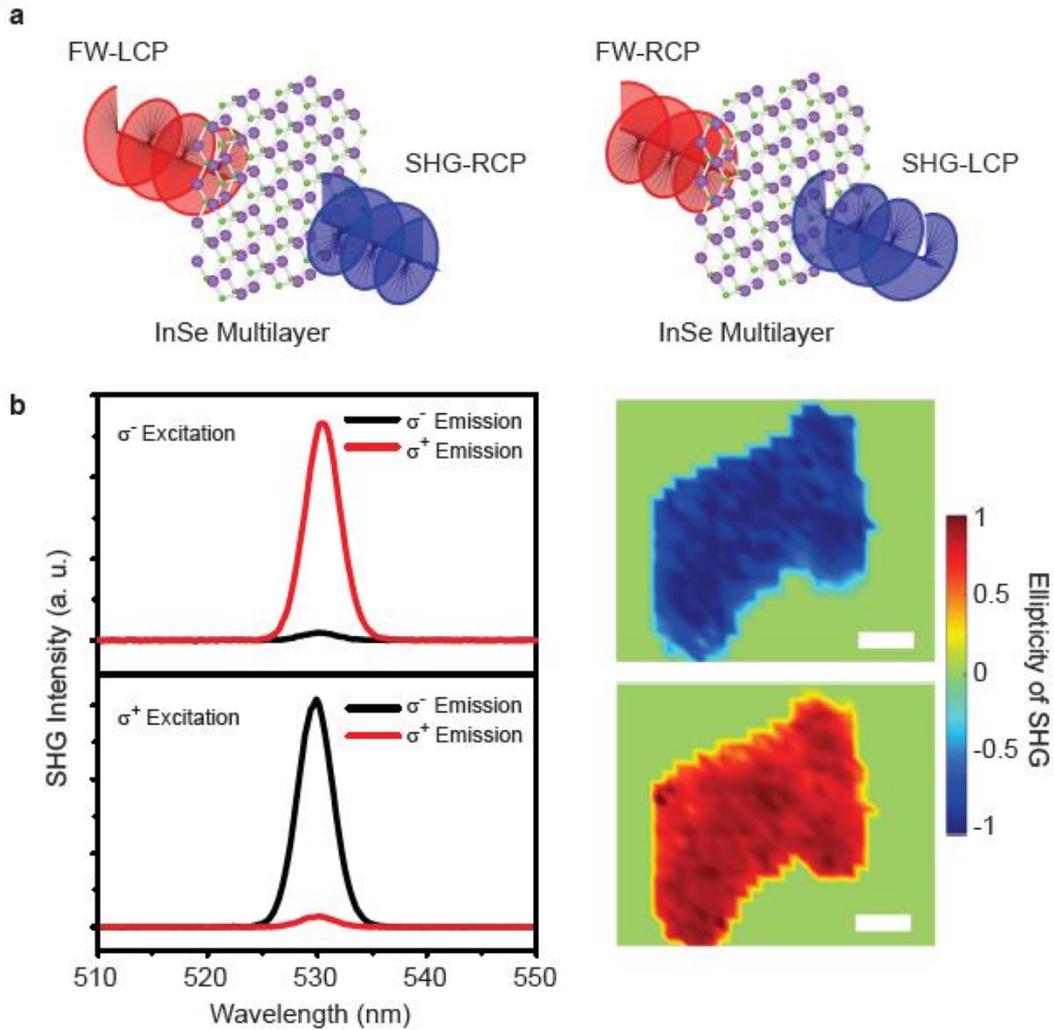

**Figure 2 Circularly polarization-resolved SHG spectra in layered InSe.** (a) Schematic of helicity-dependent SHG emission. (b) Left: circularly polarization-resolved SHG spectra excited by a left- (top) and right-circularly (bottom) polarized fundamental wave at 1060 nm, respectively. The red and black lines represent the right- and left-circular components of SHG signal at 530 nm, respectively. The SHG emission has the opposite helicity to that of the fundamental wave. Right: the ellipticity of SHG mapping excited by a left- (top) and right-circularly (bottom) polarized fundamental pulse in a layered InSe, respectively. The averaged value of ellipticity of the entire sample is around $0.93 \pm 0.05$ for LCP excitation and $-0.91 \pm 0.05$ for RCP excitation, respectively. Scale bar: 5 $\mu m$.

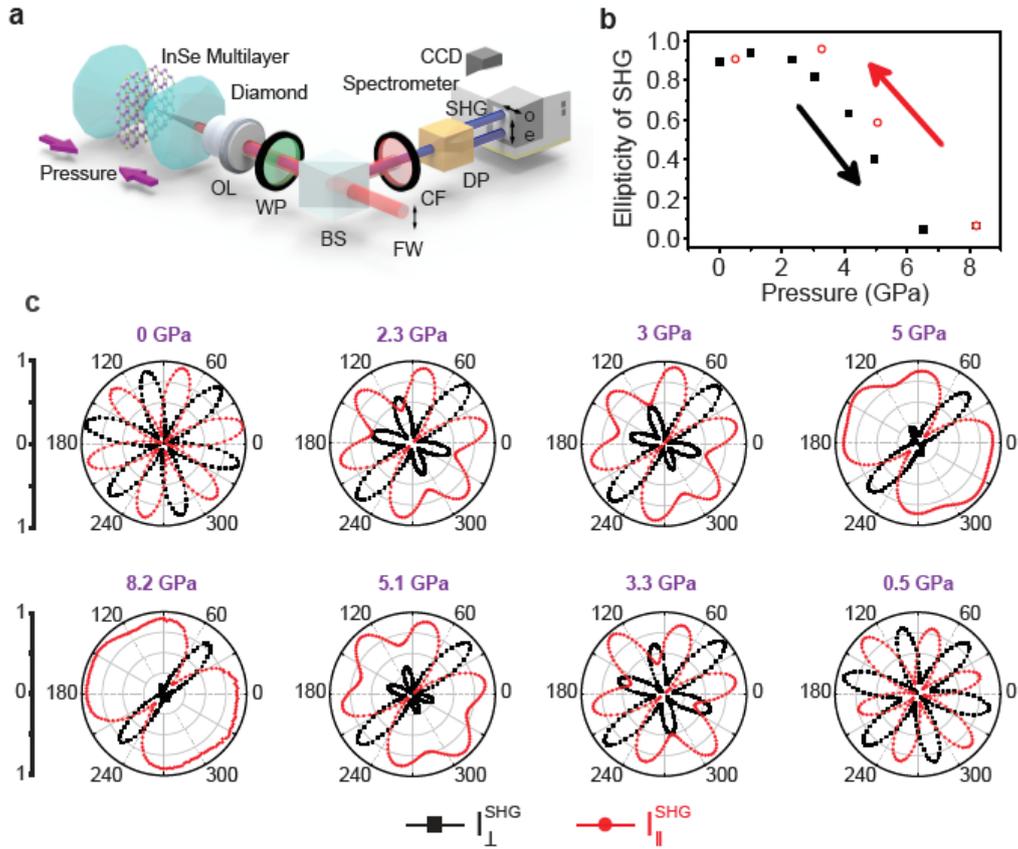

**Figure 3 Probing polarized second harmonic response of layered InSe sample under various pressures.** (a) Schematic of the polarization-resolved SHG setup under pressure, which is achieved by DAC. OL: objective lens; WP: waveplate; BS: beamsplitter; CF: color filter; DP: displacer; FW: fundamental wave. (b) The ellipticity (P) of SHG as a function of pressure in the same pressure regime. The black squares and red circles represent P in the process of increasing and decreasing pressure, respectively. (c) Eight representative polarization-resolved SHG spectra under increasing and decreasing pressures. The black squares and red circles represent the intensity of perpendicular ($I_\perp$) and parallel ($I_\parallel$) components of SHG, respectively. The pressure is increased from 0 to 8.2 GPa, and then decreased to 0.5 GPa.

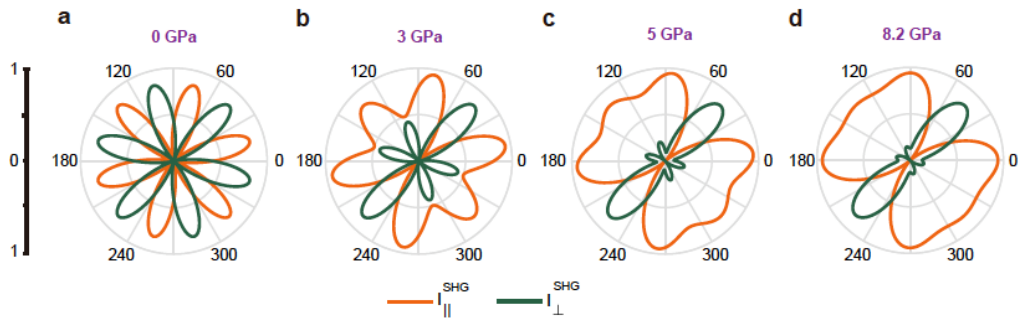

**Figure 4 Calculated polarization properties of the SHG in layered InSe under different pressures.** The green solid lines and orange solid lines correspond to the intensity of perpendicular ($I_\perp$) and parallel ($I_\parallel$) components of SHG, respectively.